\documentstyle[11pt,aaspp4]{article}
\def\mic{$\mu$m\ }
\begin{document}
\title{CAN COMPOSITE FLUFFY DUST PARTICLES SOLVE THE INTERSTELLAR CARBON
CRISIS?}
\author{Eli Dwek\altaffilmark{1}}

\altaffiltext{1}{Laboratory for Astronomy and Solar Physics, Code 685,
NASA/Goddard
Space Flight Center, Greenbelt, MD 20771.\ \ e-mail address:\
eli.dwek@gsfc.nasa.gov}

\begin{abstract}
 Recent observations suggest that the abundance of carbon in the interstellar
medium is only $\sim 60$\% of its solar value, and that other heavy elements may
be depleted by a similar amount.  Furthermore, more than half of the
interstellar carbon is observed to be in the gas in the form of C$^+$, leaving
less than $\sim 40$\% of the solar carbon abundance available for the dust
phase.  These observations have created the so-caled interstellar "carbon
crisis", since traditional interstellar dust models require about twice that
value to be tied up in graphite grains in order to explain the interstellar
extinction curve.

Recently, Mathis (1996) suggested a possible solution to this crisis. In his
newly-proposed dust model the majority of the interstellar carbon is contained
in composite and fluffy grains that are made up of silicates and amorphous
carbon grains, with $45\%$ of their volume consisting of vacuum. Per unit mass,
these grains produce more UV extinction, and can therefore account for the 
interstellar
extinction curve with about half the carbon required in traditional dust models.

This paper presents an detailed assessment of the newly-proposed dust model, and
concludes that it falls short in solving the carbon crisis, in providing a fit
to the UV-optical interstellar extinction curve.  It also predicts a
far-infrared emissivity in excess of that observed with the {\it COBE}/DIRBE and
FIRAS instruments from the diffuse interstellar medium.  This excess infrared
emission is a direct consequence of the lower albedo of the composite fluffy
dust particles, compared to that of the traditional MRN mixture of bare silicate
and graphite grains.  The failure of the new model highlights the
interrelationships between the various dust properties and their observational
consequences, and the need to satisfy them all simultaneously in any
comprehensive interstellar dust model.

In light of these problems, the paper examines other possible solutions to the 
carbon crisis.

\end{abstract}


\section{INTRODUCTION} 
The most commonly used interstellar dust model, first introduced
by Mathis, Rumpl \& Nordsieck (1977; hereafter MRN), consists of two distinct
populations of bare silicate and graphite particles with an $a^{-3.5}$ power law
distribution in grain sizes extending approximately from 0.0050 \mic to 0.25 
\mic.  With
the optical constants of Draine \& Lee (1984) and Draine (1985) the model is 
very
succesful in reproducing the average intestellar extinction curve, the 9.7 and 
18 \mic
silicate extinction features, and the average interstellar albedo and 
polarization.  The
model requires that about 75\% of the solar abundance of carbon be locked up in 
graphite
grains, and essentially all the solar abundance of Mg, Si, and Fe and about 12\% 
of the
solar abundance of oxygen be locked up silicate dust (assuming a silicate 
composition of
\{MgSiFe\}O$_4$).

There have recently been various lines of evidence for the case that the 
abundances of
various heavy elements in the interstellar medium (ISM) gas are less than their 
nominal
solar values.  Concentrating on carbon, its solar abundance, normalized to a 
hydrogen
abundance of 10$^6$ H-atoms, is $Z_{C,\odot}\approx$ 355 (Grevesse, Noels, \& 
Sauval
1996).  In the ISM, the abundance of this element is significantly lower, only
225$\pm$50 C-atoms (\cite{sw95}).  This value is based on observations of young 
stars
whose photospheric carbon abundances reflect those of the interstellar medium 
(gas plus
dust phases) at the time of their formation.  A significant amount of this 
interstellar
carbon abundance is in the gas phase.  Using the Goddard High Resolution 
Spectrograph on
board the {\it Hubble Space Telescope}, Cardelli et al.  (1996) inferred a C 
abundance
of 140$\pm$20 atoms from C II] $\lambda 2325$ \AA\ absorption measurements in 
the
direction of six stars.  This leaves a total of 85$\pm$55 C-atoms available for 
the dust
phase, compared to the $\sim$ 300 atoms required by the MRN and various other 
dust
models to be in solids.  This has created what Kim \& Martin (1996) termed the 
"C/H
crisis".  Furthermore, there is evidence that the ISM abundances of other heavy 
element,
such as nitrogen and oxygen, are depleted by as much as 60-70\% compared to 
their solar
value (see review by Mathis 1996).  The crisis is therefore not confined to 
carbon 
alone, since dust models use up
essentially all the available solar abundance of refractory elements such as Mg, 
Si, Ca,
Ti, and Fe.

This carbon crisis has motivated Mathis (1996) to propose a new interstellar 
dust model
in which most of the dust is bonded together in the form of a loose aggregate of 
dust
grains.  A previous version of this model, and the motivations behind it were 
presented
by Mathis \& Whiffen (1989).  Simply put, these fluffy aggregates can produce 
more
extinction per unit mass than their combined individual constituent dust grains. 
 An
interstellar dust model with fluffy aggregates as a major dust component will 
therefore
require a minimal amount of carbon and silicate dust, and perhaps be consistent 
with
current interstellar abundances constraints.

In addition to the abundance constraints, any interstellar dust model should be 
able to
explain the observed interstellar extinction, albedo, polarization, and infrared 
(IR)
emission.  In this paper we concentrate on the abundance, extinction, albedo, 
and IR
emission constraints placed on the composite fluffy dust (CFD) model presented 
by
Mathis.  The detection of 3.5 - 1000 \mic emission from high lattitude cirrus 
clouds by
the Diffuse Infrared Background Experiment (DIRBE) and the Far Infrared Absolute
Spectrophotometer (FIRAS) instruments on board the {\it COBE} satellite (Bernard 
et al.
1994; Weiland et al.  1996; Arendt et al.  1997, \cite{d97}) provide strong 
constraints
on interstellar dust models.  The emission in the 3.5 - 12 \mic regime can be 
used to
derive the abundance of the carriers of the 3.3, 6.2, 7.7, 8.6, and 11.3 \mic 
features,
commonly referred to as the unidentified infrared emission bands (UIBs).  The 
amount of
carbon in these carriers has to be added to any interstellar dust model.  The 
{\it COBE}
observations also provide an estimate of the fraction of the total IR radiation 
that is
emitted by these particles.  They must therefore be responsible for a 
significant
fraction of the integrated UV-optical extinction.  Finally, changes in the 
UV-optical
properties of the dust can alter their IR emissivities and temperatures, effects 
that
can be detected in the observed spectrum of the diffuse ISM.


\section{THE OPTICAL PROPERTIES OF COMPOSITE FLUFFY DUST (CFD) PARTICLES}

The new Mathis model consists of very small graphite grains which are required 
to produce
the {2200 \AA}\ extinction feature, very small silicates, and composite 
particles
consisting of loosely packed amorphous carbon and silicates particles.  In order 
to
minimize the amount of carbon needed to be in the solid phase, the model uses 
only 55
C-atoms in the form of graphite, just enough to produce the 2200 \AA\ bump.  The
remaining model parameters are then varied in order to produce the best fit to 
the
interstellar extinction curve of Cardelli, Clayton, \& Mathis (1989) with the 
minimum
amount of carbon and silicate dust.  For amorphous carbon, Mathis found that the 
samples
of type Be (see Rouleau \& Martin 1991) provided the best fit to the extinction 
data.
The silicate dust is assumed to be composed of \{MgSiFe\}O$_4$ with Ca, Al, and 
Ti
oxides as additional minor constitutents.  We characterize the silicate 
abundance by the
number of silicate atoms locked up in the dust, with the understanding that a 
comparable
number of Mg- and Fe-atoms, and about 4 times as many O- atoms are locked up in 
the dust
as well.

We reproduced the Mathis CFD model with dust particles with the following
characteristics (abundances quoted are normalized to 10$^6$ H-atoms):  (1) 
graphite
grains with a C abundance of 55 ; (2) silicate particles with a Si abundance of 
6.5; and
(3) composite particles containing 105 C-atoms in amorphous form, 26 Si-atoms in
silicates, and vacuum.  The total dust phase abundance of silicate is thus 32.5, 
about
85-92\% of its solar abundance of $\sim$ 35.5-38.  The fractional volume 
occupied by
these three composite grain constituents are 17, 38, and 45\%, respectively.  
The mass
density of the three grain constituents are 2.25 g/cm$^3$ for the graphite 
particles;
3.3 g cm$^{-3}$ for the silicates; and, adopting a value of 1.85 g cm$^{-3}$ for 
the
amorphous carbon, 1.55 g cm$^{-3}$ for the composite fluffy particles. 
Interstellar 
grains are not spherical. Consequently, we followed Mathis' procedure, and 
multiplied 
the extinction calculated for the spherical bare silicates and the composite 
particles 
by a factor of 1.09, to account for the increased extinction due to the 
oblateness of 
the dust particles. 

The dielectric properties of composite particles can be derived by averaging 
those of
their individual dust constituents.  Two of the most common rules used to derive 
the
dielectric properties of inhomogeneous particles are the Maxwell-Garnett, and 
the
Bruggemann rule (see Bohren \& Huffman 1983).  The former rule regards the 
composite
particle as a matrix with embedded inclusions, and is not invariant under the
interchange of matrix and inclusions.  The Bruggeman rule is, and for a 
composite
particle consisting of three constituents (amorphous carbon, silicates, and 
vacuum), the
average dielectric constant $\epsilon_{av}$ is given by the solution of a cubic 
equation
(see Bohren \& Huffman eq.  [8.51]):

\begin{equation} \sum^3_{j=1} {f_j {{\epsilon_j-\epsilon_{av}}\over
{\epsilon_j+2\epsilon_{av}}} } 
\end{equation}
where $\epsilon_j$ is the
dielectric constant of the j$^{th}$ constituent, and where the sum of the volume 
filling
factors, $f_j$, of the constituent material is unity.  Figure 1 compares the 
dielectric
constants of the composite particle with those of its constituents as a function 
of
wavelength.  The dielectric constants of the silicate grains were taken from 
Draine \&
Lee (1984), and those for amorphous carbon from Rouleau \& Martin (1991).

The size distribution of the composite fluffy particles is given by (Mathis 
1996):
\begin{equation} f(a)=a^{-\alpha_0}\ exp[-(\alpha_1a + \alpha_2/a +
\alpha_3 a^2)] 
\end{equation}
where $a$ is the grain radius, and the
parameters \{$\alpha_0, \alpha_1, \alpha_2, \alpha_3$\} are equal to \{3.5, 
0.0033,
0.437, 50\}, respectively (Mathis 1996, private communication).  The size 
distribution
is concentrated around 0.1 \mic, dropping to 10\% of its peak value at $a\approx 
0.05$,
and $0.2\mu$m.

Two factors contribute to the enhancement in the extinction of the composite 
grains at
infrared wavelength.  This can be seen from the extinction optical depth at 
wavelength
$\lambda$ per H-column density which is given by:
\begin{equation}
\begin{array}{lll}
\tau_{ext}(\lambda)/N_H & = & {\cal M}_d\ <\kappa_d(\lambda)> \\
                        & = & ({\cal M}_d/\rho) <3Q_{ext}/4a>
\end{array} 
\end{equation}
where ${\cal M}_d$ is the dust mass column density, and 
$<\kappa_d(\lambda)>\equiv
<3Q_{ext}/4\rho a>$ is the size-averaged dust mass extinction coefficient at 
wavelength
$\lambda$, where Q$_{ext}$ is the extinction efficiency of a grain of radius $a$ 
and
mass density $\rho$.  The mass column density of the composite grains is equal 
to the
sum of the mass column densities of its grain constituents.  However, because of 
its
porosity its mass density, $\rho$, is smaller than theirs.  As a result, the 
extinction of the
composite grains is larger than the sum of the extinction of its consituent 
particles,
which is the main reason for the enhanced extinction at UV-optical wavelengths.  
At long
wavelengths ($\lambda\gtrsim 1\ \mu$m), an additional factor plays a role in 
increasing
the extinction of the composite particles, namely an increase in their value of
Q$_{ext}$ compared to that of the amorphous carbon and silicate grains.  In the 
Rayleigh
limit, when $2\pi a/\lambda \ll\ 1$, Q$_{ext}/a$ can be written in terms of
$\epsilon_1$, and $\epsilon_2$, the real and imaginary parts of the dielectric 
constant
as:
\begin{equation}
Q_{ext}/a = {24\pi\over\lambda} {\epsilon_2 \over (\epsilon_1+2)^2+\epsilon_2^2}
\end{equation}
The value of Q$_{ext}/a\rightarrow 0$ when either $\epsilon_2\ll \epsilon_1$, or
$\epsilon_2\gg 1$.  It has a maximum when $\epsilon_2\approx\epsilon_1+2$.  
Figure 1
shows that at long wavelengths ($\lambda \gtrsim \ 100 \mu$m), $\epsilon_2 \ll
\epsilon_1$ for silicates, and $\epsilon_2 \gg 10$ for amorphous carbon.  For 
the
composite grains $\epsilon_2\approx \epsilon_1+2\ \approx 10 $, yielding a 
larger value
of Q$_{ext}$/a over that of amorphous carbon or silicate grains.  So at far-IR
wavelengths, the averaging of the optical properties plays an important role in
increasing the absorptivity/emissivity of the composite grains. As a result, the 
composite fluffy grains are somewhat cooler than either graphite or silicate 
grains of identical radius.

\section{THE EFFECTS OF THE INCLUSION OF PAHs ON THE COMPOSITE FLUFFY DUST 
MODEL}
\subsection{The Infrared Emission}

An important test for the validity of the CFD model is if it can reproduce the
observed {\it COBE}/DIRBE and FIRAS 3.5 - 1000 \mic emission from the diffuse
ISM (\cite{ber94}, \cite{was96}, \cite{ar97}, \cite{d97}).  The Mathis model
does not include carriers of the UIBs which are necessary to account for the
emission in the 3.5, 4.9, and 12 \mic DIRBE bands (\cite{d97}).  So a priori,
the CFD model needs to be ammended to include UIB carriers.  For calculational
purposes we identified these carriers with polycyclic aromatic hydrocarbons
(PAHs), with extinctions and IR properties as given by D\'esert et al.  (1990).
With this PAH model, most of the 12 $\mu$m and all shorter wavelength diffuse
emission is produced by these particles.  The abundance of PAHs can therefore be
directly determined from the near-IR {\it COBE} observations of the diffuse ISM
(see Dwek et al.  1997 for details of the model).  The PAH abundance required to
reproduce the {\it COBE} data was added to the CFD model.

We note here that whereas the diffuse ISM spectrum obtained by the {\it Infrared
Astronomical Satellite} ({\it IRAS}) could be reproduced by a simple extension
of the graphite or silicate grain size distribution to very small sizes (Draine
\& Anderson 1985; Weiland et al.  1986), the DIRBE spectrum {\it cannot} be
fitted by the same method.  The DIRBE extends the {\it IRAS} 12, 25, 60, and 100
\mic observations to shorter (and longer) wavelengths, and the near-IR 3.5 and
4.5 \mic colors observed by DIRBE from the diffuse ISM are inconsistent with
those calculated for stochastically-heated graphite or silicate grains (see Dwek
et al. 1997; Figure 2).  They are consistent, however, with those produced by
PAHs.  The addition of PAHs is therefore necessary in order to reproduce the
short wavelength diffuse IR emission observed by the {\it COBE}.

Calculation of the IR emission requires knowledge of the composition,
abundances, and size distribution of the dust particles, and the spectrum and
intensity of the ambient radiation field.  In the CFD model, all the dust
parameters are determined by the fit to the interstellar extinction curve,
except for the size distribution of the bare silicate and graphite grains.
Their size distribution was not specified by Mathis (1996) since it does not
affect the extinction as long as the dust grains are Rayleigh particles.  For
the purpose of our calculations we adopted an MRN power law distribution,
$dn(a)/da\sim a^{-3.5}$ for the bare silicate and graphite grains in the 0.005
to 0.015 \mic radius interval.  The upper limit corresponds to the largest grain
radius for which the value of $Q_{ext}/a$ is still independent of grain size.

Dust in the diffuse high latitude clouds is heated by the local interstellar
radiation field (LISRF), and for calculational purposes we adopted the LISRF
spectrum and intensity from the studies of Mathis, Mezger, \& Panagia (1983).
With the dust model and interstellar radiation field specified, we calculated
the emerging IR emission without further adjustments of any model parameters.
Our calculations include the effects of the stochastic heating and temperature
fluctuations on the IR dust spectrum (see Dwek et al.  1997 for details).

Figure 3 compares the IR emission predicted by the CFD model to the
observations.  With the addition of PAHs, the model gives a very good fit to the
observed intensity at $\lambda \lesssim 140\ \mu$m.  PAHs radiate about 25\% of
the total diffuse IR emission (Dwek et al.  1997).  However, at longer
wavelengths the model produces a large excess of emission over that detected by
the {\it COBE} satellite.  As shown below, the enhanced IR emission is an
unavoidable consequence of the UV-optical properties of the CFD particles.

\subsection{The Interstellar Extinction and Dust Albedo}

Figure 4 presents the extinction predicted by the CFD model in the various
wavelength regimes.  Figure 4a depicts the UV-optical portion of the curve as a
function of $x \equiv 1/\lambda(\mu m)$.  The thin line in the figure shows the
extinction presented by Mathis (1996) without PAHs.  The model provides a very
good fit to the data, deviating by less than 10\% over the given range of $x$.
The fit would have been better if, following Mathis (1996), we had modified the
optical constants of the silicate grains in the 6 - 8 $\mu$m$^{-1}$ region to
elliminate an artificial kink in their extinction curve at $\sim 7\
\mu$m$^{-1}$.  The bold line in the figure depicts the total extinction when
PAHs are included in the dust model.  The calculated extinction is now
significantly higher than the observed one, which should not be surprising,
since PAHs radiate about 25\% of the total diffuse IR emission.  The {\it shape}
of the PAH extinction is highly uncertain.  Here we adopted the extinction
properties of the PAHs from the work of D\'esert et al.  (1990), which were
chosen to reproduce the general interstellar extinction curve with their dust
model.  However, regardless of the exact form of their UV-optical cross
sections, PAHs (or any other carriers of the UIBs) must be responsible for about
$25(1-A)$\% of the integrated extinction in the $1/\lambda=1 - 10\ \mu m^{-1}$
interval, where $A$ is the average albedo at UV-optical wavelength.

Figure 4b compares the extinction produced by the CFD model (including PAHs) to
the observed infrared extinction (Mathis 1990; Table 1).  The model produces
higher extinction at wavelengths $\gtrsim 10 \mu$m, but it is nevertheless
consistent with the observations considering the fact that uncertainties in the
data are at least a factor of two at $\lambda \gtrsim 15 \mu$m.

Figure 5 depicts the albedo of the CFD model as a function of $x \equiv
1/\lambda(\mu m)$ (bold line).  For comparison, we also plotted the albedo
predicted by the MRN model with Draine-Lee optical constants (dotted line).  In
particular, the visual albedo of the MRN model is about 0.6, consistent with the
value of $0.61\pm 0.07$ suggested by Witt (1989).  The albedo of the CFD model
is lower, about $0.5$, a fact already pointed out by Mathis (1996).  So in spite
of the fact that both, the MRN and the CFD, models reproduce the observed
interstellar extinction equally well, the CFD model has a lower albedo over the
whole UV-optical wavelength range.  The composite fluffy dust particles
therefore absorb more energy from the LISRF than the bare silicate graphite
grains in the MRN model.  This energy has to be reradiated at IR wavelengths,
producing an excess far-IR emission compared to that produced by an MRN
distribution of bare graphite and silicate grains.

\subsection{The Carbon Abundance}

The main motivation for the composite dust model was to solve the interstellar
carbon crisis.  The original MRN model requires a C abundance of about 270 atoms
to be in the dust (using Draine-Lee optical constants), significantly above the
value of 85$\pm$55 implied from the recent ISM values.  Kim \& Martin (1996)
optimized the grain size distribution, and produced a similar C abundance of
270$\pm$50 atoms in carbon dust.  The CFD model of Mathis takes therefore a
large step towards easing the carbon crisis, requiring only $\sim$160 C-atoms to
be in the dust phase.  However, the model ignores the amount of carbon locked up
in PAHs which was recently estimated from the {\it COBE} data to be 70$\pm$20
C-atoms (\cite{d97}).  This PAH abundance is higher by about a factor of two
compared to previous estimates (e.g.  D\'esert et al.  1990; Siebenmorgen \&
Kr\"ugel 1992).  However, recent observations of the 6.05 \mic absorption
feature in the ISM with the {\it Infrared Space Observatory} ({\it ISO}; Schutte
et al.  1996) suggest a similarly large value.  Solid H$_2$O provides the bulk
of the absorption in the 6.05 \mic band.  However, an excess absorption in the
red wing of the feature is due to the C-C stretching mode, and the authors
estimate that about 20\% of the solar carbon abundance (they adopt (C/H)$_\odot
= 10^{-4}$) must be in this bond to produce the observed absorption.  If this
bond is in PAHs, then these observations confirm the PAH abundance estimates of
Dwek et al.  (1997).  If we add PAHs to the composite dust model we get that it
will require a total C abundance of 225$\pm$20 atoms (55 in small graphite
grains, 105 as amorphous carbon in composite grains, and 70 in PAHs).  This
carbon abundance is smaller than any other dust model [see Kim \& Martin (1996)
or Snow \& Witt (1995) for a summary of the carbon abundance requirements of the
various dust models], but still significantly larger that the maximum available
for the dust in the ISM.

\subsection{PAHs as Carriers of the 2200 \AA\ extinction feature?}

One way to ease the carbon abundance constraint is to assume that PAHs are the
carriers of the 2200 \AA\ extinction bump as well as the UIBs.  This assumption
will reduce the C abundance by 55 atoms, the amount locked up in graphite
grains.  There are several arguments in favor of such an identification:  (1)
Laboratory measurements show that small PAHs have an enhanced UV absorptivity
around 2200 \AA\ (Joblin, L\'eger, \& Martin 1992).  Peak cross sections are
about $8\ 10^{-18}\ cm^2$ per C-atom.  For an opacity of $\tau_{2200}/N_H=5.27\
10^{-22}\ cm^{-2}$, and a Drude profile for the emission bump (e.g.  Draine 
1989), the required C abundance in PAHs is
$\sim$ 65 atoms, comparable with the amount required to produce the UIB emission
features; (2) extinction measurements along several lines of sight through the
Chamaeleon cloud (\cite{bpg94}) show a corelation between the strength of the
2200 \AA\ bump and the magnitude of the IRAS 12 \mic emission.  Since most of
the 12 \mic emission originates from PAHs, the correlation suggests that PAHs
can be significant contributors to the UV extinction bump.

The problem with PAHs is that the peak wavelength of the UV emission bump and 
its profile
seems to vary with PAH composition.  Furthermore, the PAH cross sections exhibit 
a
feature around 3000 \AA\ (Joblin et al.  1992) that is not observed in the ISM.  
It yet
remains to be seen if a suitable mixture of PAHs will match the interstellar 
2200 \AA\
extinction profile and smooth out the unobserved UV feature.

Even if PAHs can be substituted for graphite, the elimination of these particles 
will
create an additional problem for the CFD model.  The very small graphite 
particles are
not only responsible for the 2200 \AA\ extinction, but also contribute to the 
observed 25
- 60 \mic excess emission, since they are stochastically heated by the LISRF.  A
population of very small particles will therefore be needed to account for the 
observed
mid-IR emission, which is currently absent in the size distribution of the 
composite
grains.

\section{SUMMARY}
In this paper we showed that composite interstellar dust particles
consisting of amorphous carbon and silicate dust with vacuum comprising 45\% of 
their
volume cannot be a major interstellar dust component in the diffuse ISM.  Heated 
by the
local interstellar radiation field, these particles produce a significantly 
higher IR
emission than observed by the {\it COBE}/DIRBE and FIRAS instruments for the 
diffuse high
latitude emission.  The excess IR emission is the result of the increased 
absorptivity of
the composite fluffy dust particles at UV-optical wavelengths.  The robustness 
of this
conclusion depends on the validity of the calculations of the optical properties 
of
composite dust particles.  A different rule for producing the average optical 
properties
of the composite particles may produce an IR emissivity that is consistent with 
the {\it
COBE} data, but it will certainly affect the UV-optical extinction as well, 
requiring a
whole new assesment of the composite dust model.

The neglect of the composite fluffy dust model (or any other dust model) to 
include the
carriers of the unidentified IR emission bands (UIBs) will result in an 
underestimate of
the carbon abundance by $\sim$20\% of its solar value, and will introduce a 
$f(1-A)$\%
uncertainty in the UV-optical part of the interstellar extinction curve, where 
$A\approx
0.6$ is the average UV-optical albedo of the dust, and $f\approx 0.25$ is the 
fraction of
the total IR radiation emitted by these particles.  Since the shape of the 
UV-optical
absorptivity of the coal or PAHs that are responsible for the UIBs is unknown, 
it
introduces a major uncertainty in any interstellar extinction model.

We point out that the inferred abundance of the UIB carriers is inversely 
proportional to
their wavelength-averaged UV-optical absorption cross section.  Lowering the 
amount of
carbon in these particles will require an increase in their UV-optical 
extinction, and
conversely, minimizing their effects on the extinction will require a high C 
abundance to
be tied up in these carriers.  To summarize, the composite fluffy dust model 
does not
reproduce the UV-optical extinction to better than about 12\%, and with the 
addition of
UIB carriers it uses up twice the amount of the carbon believed to be available 
in the
ISM for the dust.

Since the composite dust model falls short in solving the carbon crisis it is
worth examining other possible solutions to this crisis.  The major assumption
leading to this crisis is that the dust abundance should reflects that of the
current local ISM.  This assumption requires that the ISM gas and dust have
identical spatial and temporal histories.  However, there is evidence to suggest
that this may not be the case:  (a) observations of a large sample of stars in
the solar neighborhood (e.g.  Edvardsson et al.  1993) shows an intrinsic
scatter in their observed abundances as a function of time and metallicity; (b)
the presence of isotopic anomalies in meteorites suggests that the solar
neighborhood may have been polluted by the nucleosynthetic products of a
neigboring stars and supernovae (e.g.  Zinner 1996); (c) additional
inhomogeneities may be introduced by dynamical fractionation between gas and
dust in star formation processes.  Numerical simulations of protostellar
collapses (Ciolek \& Mouschovias 1996) show that ambipolar diffusion can
"strain" the infalling material, leaving the grains behind and reducing their
abundance in the collapsed core which is destined to become a star.

We conclude that inhomogeneous chemical mixing and fractionation effects in the
ISM may be responsible for the differences in metallicities between the sun and
the local ISM, and perhaps between that inferred from the B stars and the
extinction as well.

\acknowledgments

I acknowledge the Institut d'Astrophysique Spatiale in Orsay, and the Institut
d'Astrophysique de Paris for providing a stimulating atmosphere and environment
for the duration of my research.  During my work I benefitted from many
stimulating discussions with Francois Boulanger, Xavier D\'esert, Ant Jones,
Alain L\'eger, Alain Omont, Renauld Papoular, Jean-Loup Puget, and Bill Reach.
Special thanks are due to John Mathis who provided stimulating discussions, some
intermediate results of his model which were used to check current calculations,
and comments on a preliminary version of the manuscript.  Comments by an
anonymous referee led to further improvement in the manuscript.  Finally, I
acknowledge the NASA/GSFC Research and Study Fellowship Program for providing
financial assistance during my stay at these institutes.


\clearpage


\begin{figure} 
\caption{(a) The values of the real part of the dielectric constant of
amorphous carbon of type Be (dashed line) and silicate (light solid line) as a 
function
of wavelength.  The dotted line at $\epsilon$(real) = 1.0 represents the vacuum. 
 The
bold line is the value of $\epsilon$(real) of the composite fluffy grain 
obtained from
equation (1).  (b) The same as (a) for the imaginary part of the dielectric 
constant.}
\end{figure}

\begin{figure} 
\caption{The value of Q$_{ext}$/a in the Rayleigh approximation (see eq.
(3) in the text) is plotted as a function of wavelength for Be-type amorphous 
carbon of
type Be (dashed line), silicate (light solid line), and the composite grains 
(bold line).
Averaging the optical constants of the constituent grains yields an enhanced 
emissivity
for the composite particle.}  
\end{figure}

\begin{figure} 
\caption{Comparison of the infrared spectrum predicted by the composite
dust model (bold line) to the average dust spectrum in the diffuse ISM observed 
by the
{\it COBE} satellite (DIRBE data are represented by diamonds, FIRAS data by the 
thin
solid line).  PAHs have been added to the model to account for the 3.5 to 12 
\mic DIRBE
emission.  Also shown in the figure is the contribution of PAHs (dotted line), 
graphite
(dashed line), silicates (dashed-dotted line), and the composite particles 
(solid line).}
\end{figure}

\begin{figure} 
\caption{The extinction of the composite dust model is compared to the
observed average interstellar extinction curve.  (a) Comparison to the 
UV-optical
extinction given by Cardelli, Mathis, \& Clayton (1989).  The addition of PAHs 
to the
composite dust model produces an excess in the extinction (bold line) over the
observations.  The thin solid line is the total extinction calculated by Mathis 
(1996)
without PAHs.  Also shown in the figure are the contributions of graphite 
(dashed line),
silicates (dotted line), and the composite grains (dashed dotted line) to the 
total
extinction.  (b) Comparison of the calculated extinction (bold line) to the 
infrared
extinction (Mathis 1990; diamonds).  The excess extinction predicetd by the 
composite
dust model is within the uncertainty of the observations.}  
\end{figure}

\begin{figure}
\caption{The albedo of the composite fluffy dust (CFD) model (solid line) is 
plotted 
versus inverse wavelength. Also shown is the albedo of the MRN model, calculated 
with 
Draine-Lee optical constants. The low albedo (high absorptivity) of the CFD 
model is the 
cause of the excess far-IR emission depicted in Figure 3.}
\end{figure}


\begin{thebibliography}{}
\bibitem[Arendt et al. 1997]{ar97} Arendt, R. G. et al. 1997, in preparation
\bibitem[Bernard et al. 1994]{ber94} Bernard, J. -P., Boulanger, F., D\'esert,
F.,-X.,
Giard, M., Helou, G.,\& Puget, J. -L. 1994, \aap, 291, L5
\bibitem[Bohren \& Huffman 1983]{bh83} Bohren, C. F., \& Huffman, D. R. 1983,
Absorption and
Scattering of Light by Small Particles (Wiley: New York)
\bibitem[Boulanger, Pr\'evot, \& Gry 1994]{bpg94} Boulanger, F., Pr\'evot, M.
L., \& Gry, C.
1994, \aap, 284, 956
\bibitem[Cardeli et al. 1989]{ccm89} Cardelli, J. A., Clayton, G. C., \& Mathis,
J. S. 1989,
\apj, 245, 345
\bibitem[Cardeli et al. 1996]{cmjs96} Cardelli, J. A., Meyer, D. M., Jura, M.,
\& Savage, B. D.
1996, \apj, 467, 334
\bibitem[Ciolek \& Mouschovias 1996]{cm96} Ciolek, G. E., \& Mouschovias, T. Ch.
1996, \apj,468, 749
\bibitem[D\'esert et al. 1990]{des90} D\'esert, F. -X., Boulanger, F., \& Puget,
J. L.1990, \aap, 237, 215
\bibitem[Draine \& Lee 1984]{dl84} Draine, B. T., \& Lee, H. M. 1984, \apj, 285,
89
bibitem[Draine \& Anderson 1985]{da85} Draine, B. T., \& Anderson, N. 1985, 
\apj, 292,
494
\bibitem[Draine 1985]{d85} Draine, B. T. 1985, \apjs, 57, 587
\bibitem[Draine 1989]{d89} Draine, B. T. 1989, in IAU Symp. No. 135,
Interstellar Dust,
eds. L. J. Allamandola \& A. G. G. M. Tielens (Boston: Kluwer), p. 313
\bibitem[Dwek et al. 1997]{d97} Dwek, E. et al. 1997, \apj, 475, 000
\bibitem[Edvardsson et al 1993]{e93} Edvardsson, B., Andersen, J., Gustafsson,
B., Lambert,
D. L., Nissen, P. E., \& Tomkin, J. 1993, \aap, 275, 101
\bibitem[Grevesse, Noels, \& Sauval 1996]{gns96} Grevesse, N., Noels, A., \
Sauval, A. J. 1996,
in Cosmic Abundances, ASP Conference Series, eds. S. S. Holts, \& G. Sonneborn
(San Francisco:
ASPCS), p. 117
\bibitem[Joblin, L\'eger, \& Martin 1992]{jlm92} Joblin, C., L\'eger, A., \&
Martin, P.
1992, \apj, 393, L79 
\bibitem[Kim \& Martin 1996]{km96} Kim, S-H, Martin, P. G. 1996, \apj, 462, 296
\bibitem[Mathis, Rumpl, \& Nordsieck 1977]{mrn77} Mathis, J. S., Rumpl, W., \&
Nordsieck,
K. H. 1977, \apj, 217, 425, MRN
\bibitem[Mathis, Mezger, \& Panagia 1983]{mmp83} Mathis, J. S., Mezger, P. G.,
\&
Panagia, N. 1983, \aap, 128, 212
\bibitem[Mathis \& Whiffen 1989]{mw89} Mathis, J. S., Whiffen, G. 1989, \apj,
341, 808
\bibitem[Mathis, J. S. 1990]{m90} Mathis, J. S. 1990, \araa, 28, 37
\bibitem[Mathis 1996]{m96} Mathis, J. S. 1996, ApJ, 472, 643
\bibitem[Rouleau \& Martin]{rm91} Rouleau, F., \& Martin, P. G. 1991, \apj, 377,
526
\bibitem[Schutte et al. 1996]{s96} Schutte, W. A. 1996, \aap, 315, L333
\bibitem[Siebenmorgen & Kr\"ugel 1992]{sk92} Siebenmorgen, R., \& Kr\"ugel, E.
1992,\aap, 259, 614
\bibitem[Snow \& Witt 1995]{sw95} Snow, T. P., \& Witt, A. N. 1995, Science,
270, 1455
\bibitem[Weiland et al. 1986]{weil86} Weiland, J. L., Blitz, L., Dwek, E., 
Hauser, M. G.,
Magnani, L.,\& Rickard, L. J. 1986, \apjl, 306, L101
\bibitem[Weiland, Arendt, \& Sodroski 1996]{was96} Weiland, J. L., Arendt, R.
G., \& Sodroski, T. J. 1996, in AIP Conf. Proc. No. 348, Unveiling the Cosmic 
Infrared
Background, ed. E. Dwek (New York: AIP), p. 74
\bibitem[Witt 1989]{w89} Witt, A. N. 1989, in Interstellar Dust, eds. L. J. 
Allamandola 
\& A. G. G. M. Tielens (Dordrecht: Kluwer), p. 87
\bibitem[Zinner 1996]{z96} Zinner, E. 1996, in Cosmic Abundances, ASP Conference
Series, eds. S. S. Holts, \& G. Sonneborn (San Francisco: ASPCS), p. 147
\end{thebibliography}
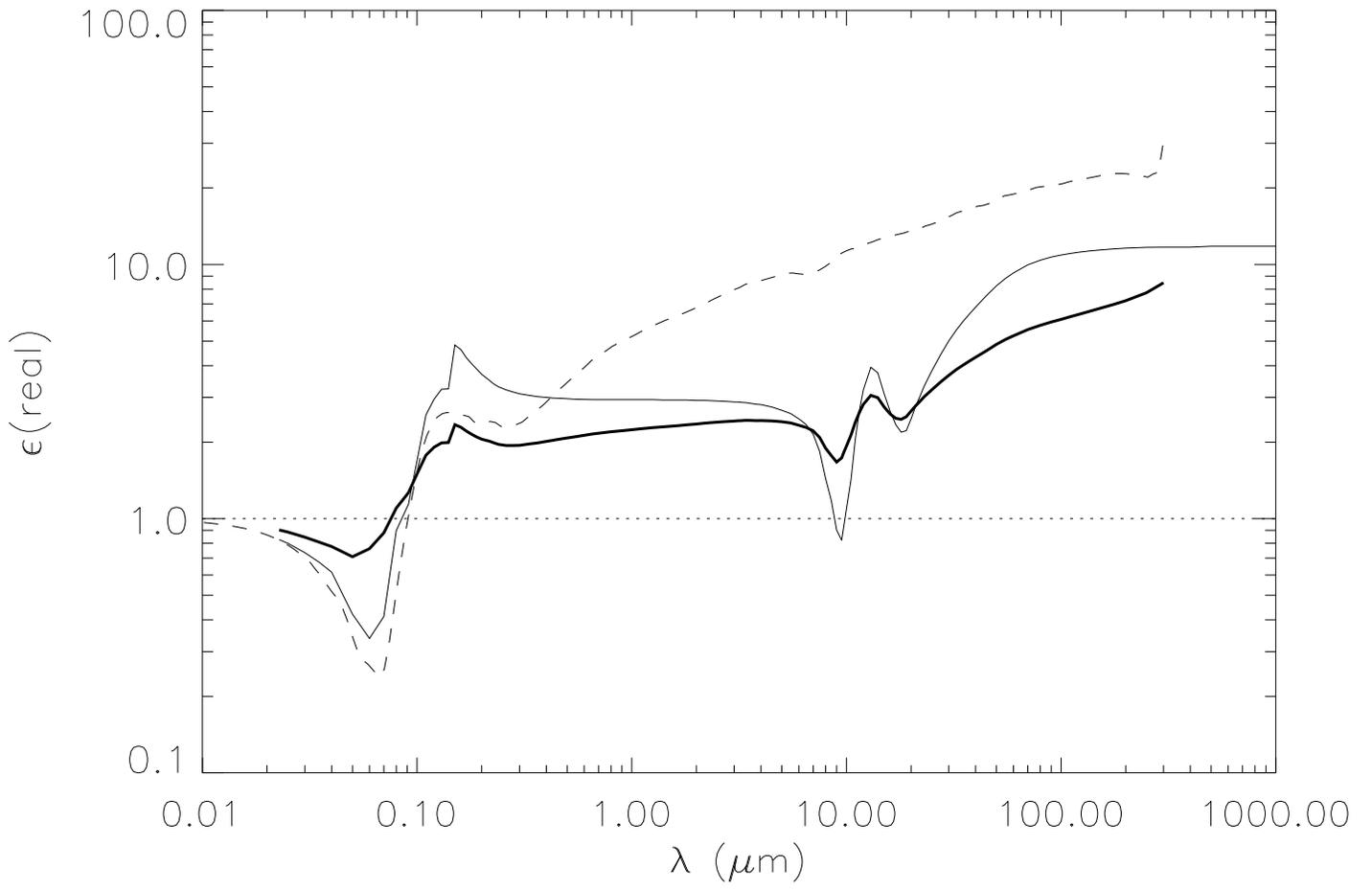
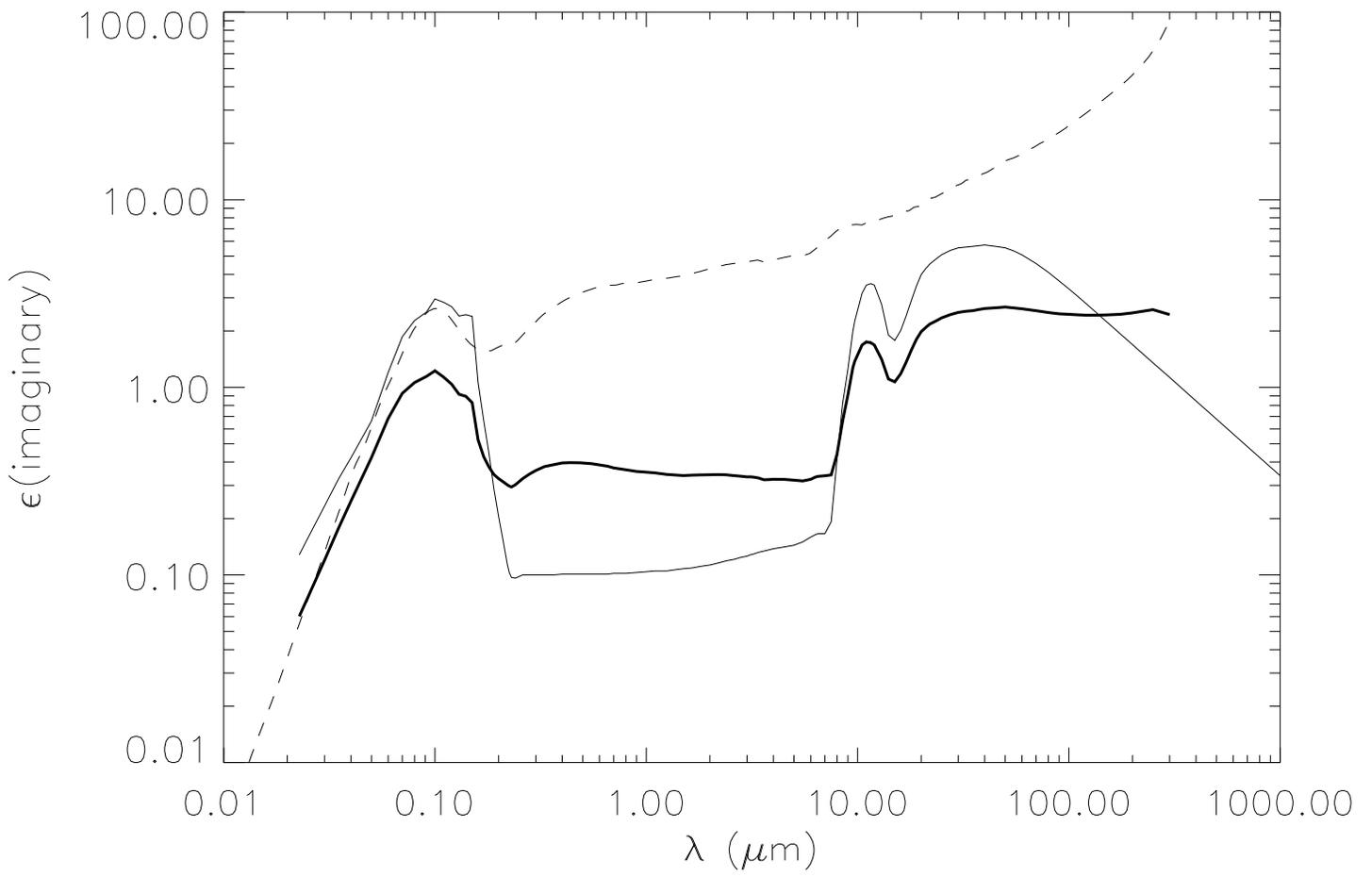
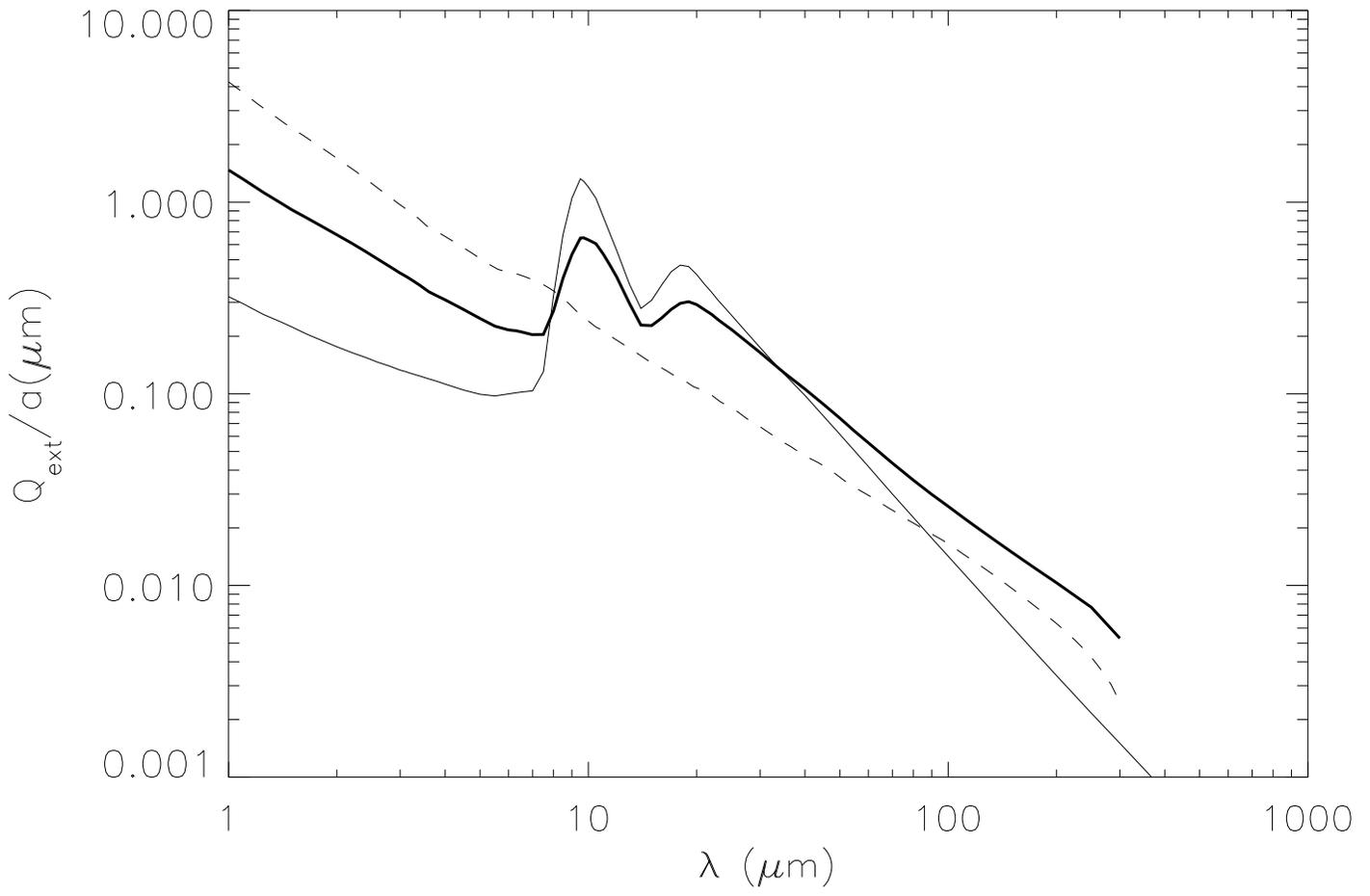
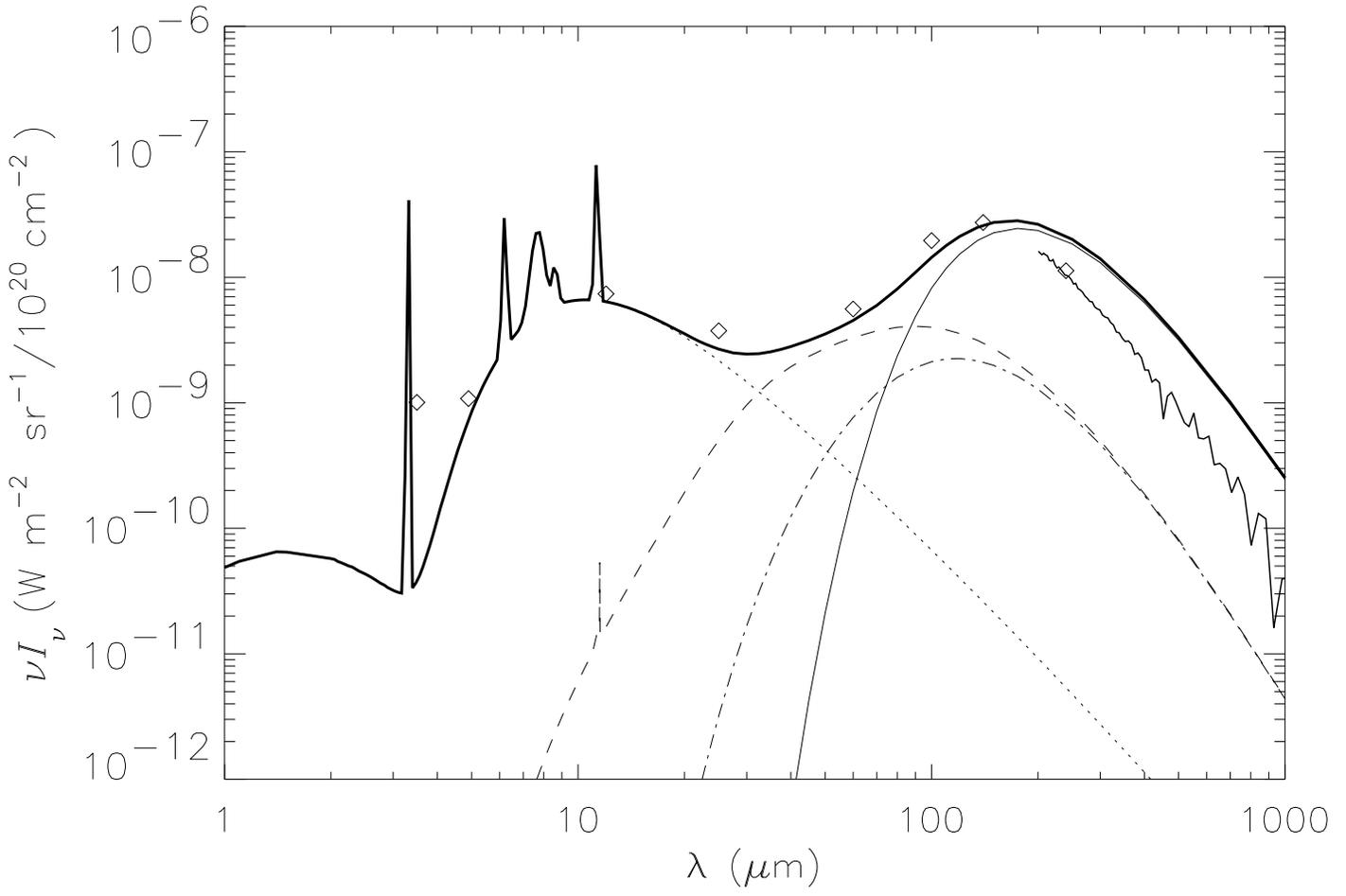
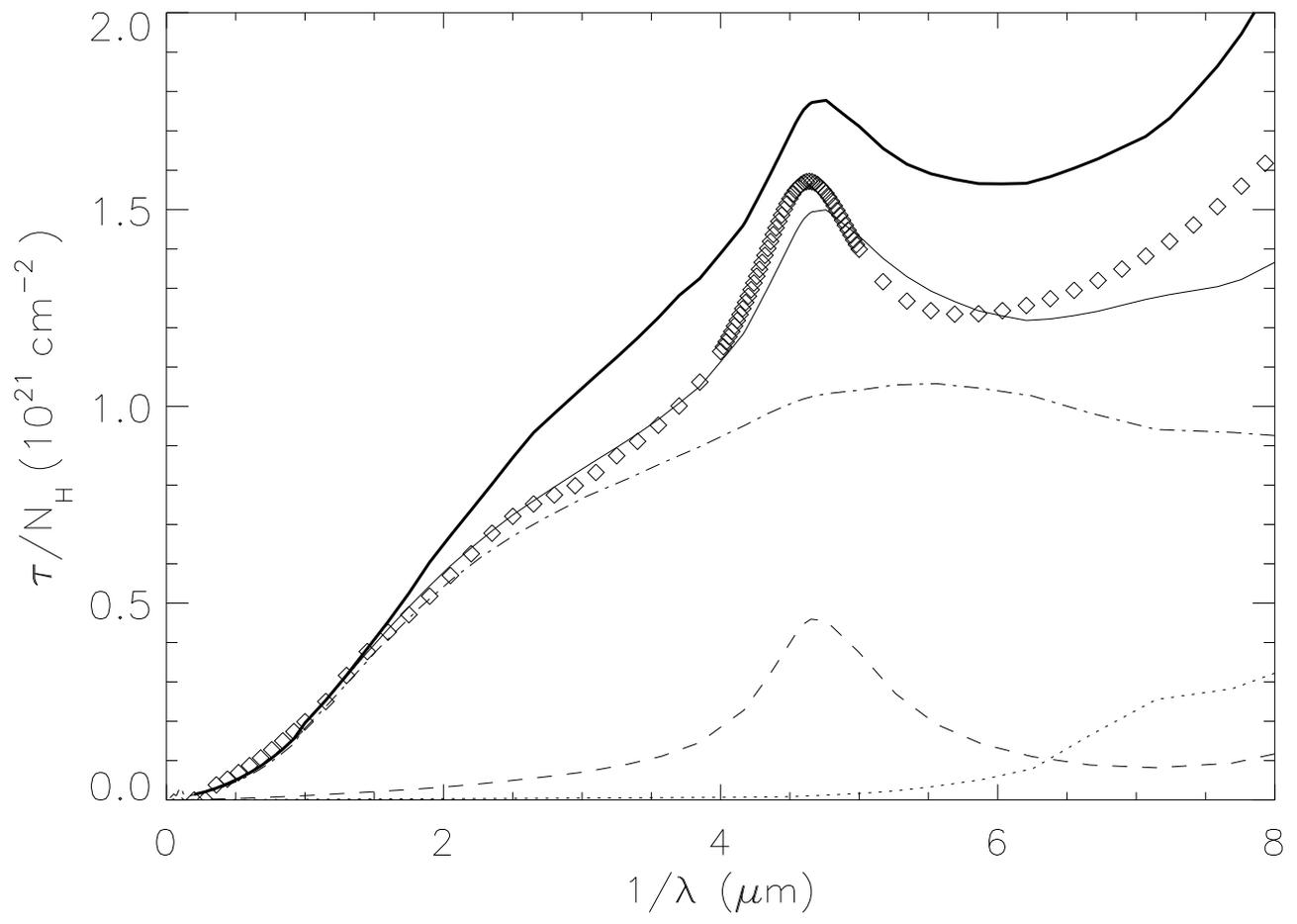
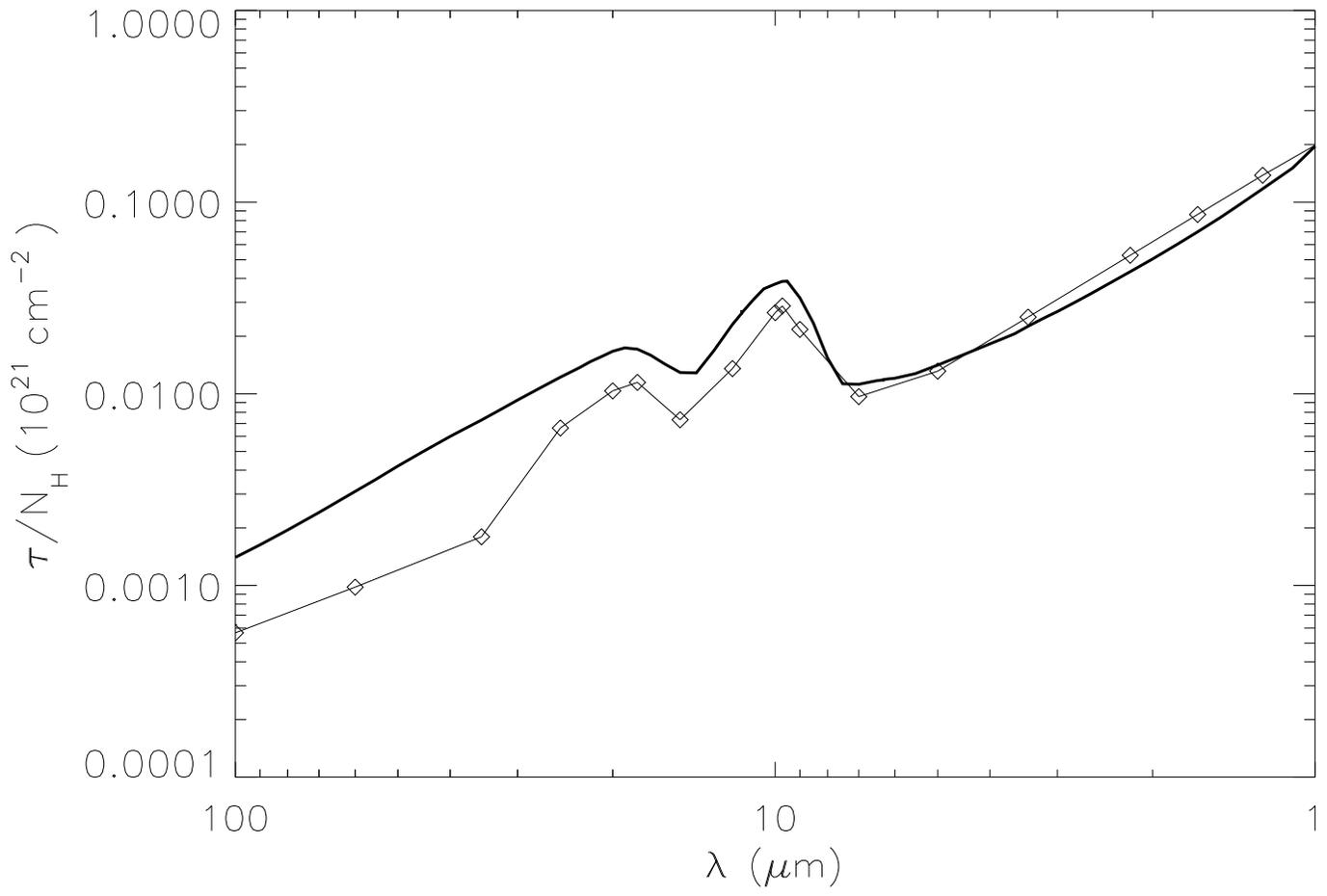
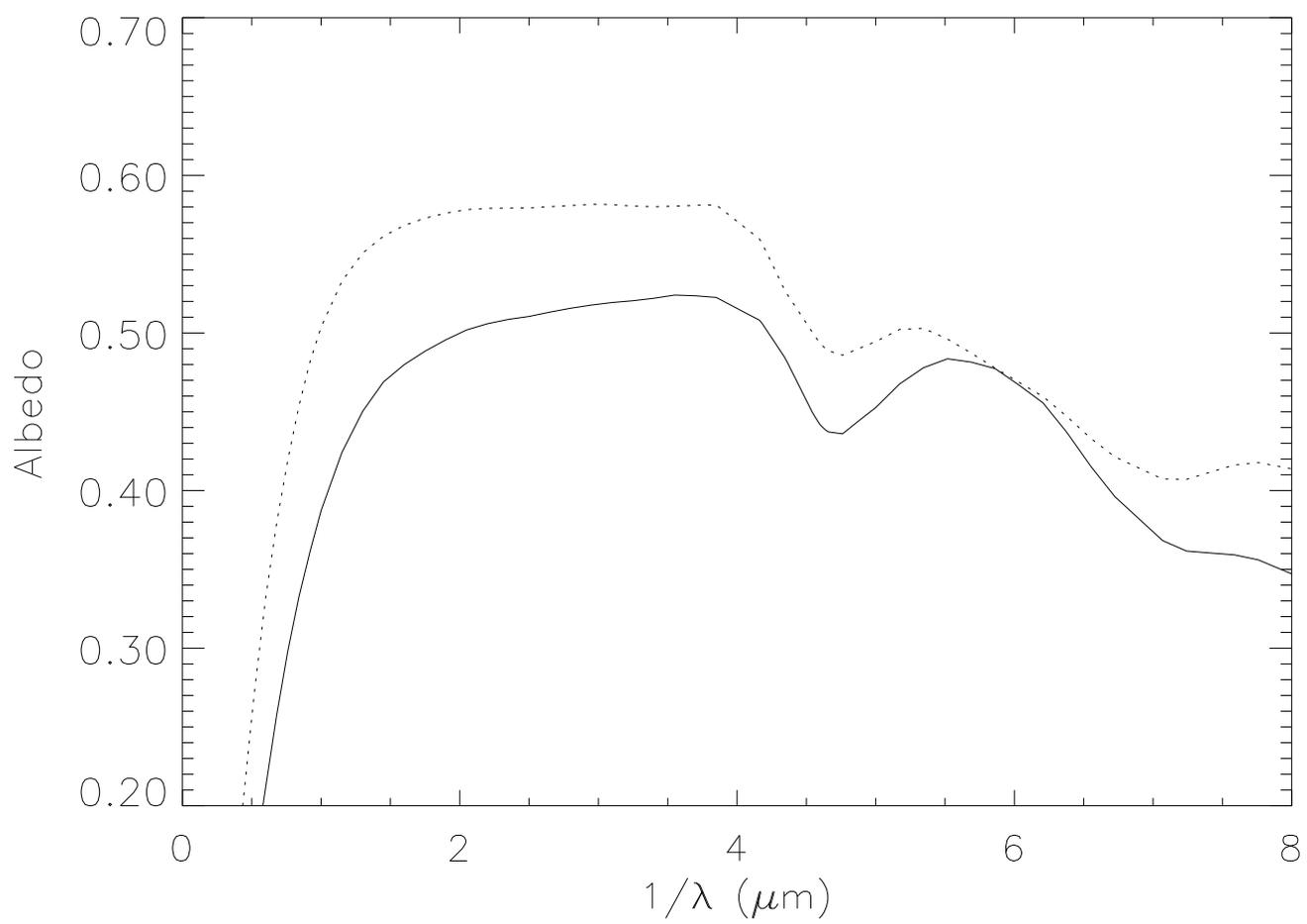
\end{document}